# Spectral perturbation and reconstructability of complex networks

D. Liu, H. Wang, and P. Van Mieghem

*Delft University of Technology, Delft, The Netherlands*
(Received 21 September 2009; published 6 January 2010)

In recent years, many network perturbation techniques, such as topological perturbations and service perturbations, were employed to study and improve the robustness of complex networks. However, there is no general way to evaluate the network robustness. In this paper, we propose a global measure for a network, the reconstructability coefficient $\theta$, defined as the maximum number of eigenvalues that can be removed, subject to the condition that the adjacency matrix can be reconstructed exactly. Our main finding is that a linear scaling law, $E[\theta]=aN$, seems universal in that it holds for all networks that we have studied.

          

## I. INTRODUCTION

"Robustness" or "resilience" properties of complex networks are commonly analyzed by perturbing the network [1,2]. The simplest perturbations, which we call elementary changes, are the omission or addition of a link and/or a node or the rewiring of links. Any topological perturbation can be constructed as a sequence of elementary changes. The degree of degradation of the network performance, measured in terms of some sets of graph metrics, under a certain topological perturbation is commonly regarded as a measure of the robustness of that network. Service perturbations are usually much more complex to define and to analyze because they consist of changes in traffic, latency, availability, and many other unknown processes. Spectral graph theory has been applied to understand properties of networks [3,4]. Meanwhile, the eigenvalue perturbation method has been well developed and applied in quantum mechanics [5,6]. Another spectral operation, the Karhunen Loéve transform (KLT), is widely used in the area of image data compression [7].

In this paper, we explore the effect of perturbing the spectra of complex networks. The eigenvalues of most complex graphs are nondegenerate and their corresponding eigenvectors are unique. As explained in Sec. II A and provided all eigenvectors are known, we remove eigenvalues from the spectrum of a graph and check whether the graph can still be exactly reconstructed by exploiting the zero-one nature of the elements of the adjacency matrix. The requirement of *exact* reconstruction is unique and in contrast to, for example, image compression, where some image information is lost. Removing eigenvalues can be regarded as a particular type of spectral perturbation. On the other hand, the eigenvectors can also be perturbed by, for example, adding noise to each component of the eigenvectors. One could as well consider a metric such as the connectivity, hop count, diameter, clustering coefficient, etc., and investigate its degradation under such spectral perturbation (eigenvalue removal and noise addition to eigenvectors).

A large number of topological measures has been derived (see, e.g., [8]) from the adjacency matrix and Laplacian to capture different features of a network as well as to classify networks. Examples are the distribution of the nodal degree, which is the number of links incident to a node, and the average clustering coefficient, which describes the link density among the direct neighbors of a node. These measures can be highly correlated, pointing to a certain level of redundancy among them. It is highly desirable to determine a specific small set of independent measures that is sufficient to characterize the network structure. However, the questions, "which measures possess more information regarding the topology" and "what is the level of redundancy or correlation between measures" are far from understood.

In this paper, we examine the maximal number of eigenvalues that can be removed from the spectrum given the set of eigenvectors. The method and related theory are explained in Sec. II. We define the reconstructability coefficient $\theta$ of a network as the maximum number of eigenvalues that can be set to zero, given that the adjacency matrix can be reconstructed exactly. Via extensive simulations on different classes of graph, presented in Secs. III and IV, we found the remarkable linear scaling law

$$E[\theta] = aN, \qquad (1)$$

where the real number $a \in [0,1]$ depends on the graph $G$. Moreover, the variance $\mathrm{Var}[\theta]$ was sufficiently smaller than the mean $E[\theta]$ such that $E[\theta]$ serves as an excellent estimate for $\theta$. For sufficiently large $N$, the law (1) tells us that a portion $a$ of the smallest eigenvalues (in absolute value) can be ignored or removed from the spectrum and that the adjacency matrix is still reconstructable (provided we have the exact eigenvectors). Section V investigates the sensitivity of the reconstructability coefficient $\theta$ under random perturbation of the eigenvectors in the class of Erdős-Rényi random graphs and concludes that the perturbation can be as high as 10% (the norm of an eigenvector is 1). The reconstructability coefficient $\theta$ [or the scaled one coefficient $a = \frac{E[\theta]}{N}$ in Eq. (1)] can be regarded as a spectral metrics of the graph that expresses how many dimensions of the $N$-dimensional space are needed to represent or reconstruct the graph. Roughly, a high reconstructability coefficient $\theta$ reflects a "geometrically simple" graph that only needs a few orthogonal dimensions to be described. The precise physical or topological meaning of the reconstructability coefficient $\theta$ is not yet entirely clear.

## II. SPECTRAL PERTURBATION

### A. Description and definition of reconstructability

The topology of a network $G$ consisting of $N$ nodes and $L$ links can be described by the adjacency matrix $A$, a $N \times N$

zero-one matrix, where the element $a_{ij} = 1$ if there is a link between node $i$ and node $j$, else $a_{ij} = 0$. Assuming that the graph is undirected, the adjacency matrix $A$ is symmetric. All eigenvalues are real and $A$ possess an eigenvalue decomposition [3]

$$A = X\Lambda X^T,$$

where $X = [x_1 \; x_2 \; \cdots \; x_N]$ is an orthogonal matrix (such that $X^T X = X X^T = I$) with as columns the real and normalized eigenvectors $x_1, x_2, \ldots, x_N$ of $A$, corresponding to the eigenvalues $\lambda_1 \geq \lambda_2 \geq \cdots \geq \lambda_{N-1} \geq \lambda_N$ in descending order and the diagonal matrix $\Lambda = \text{diag}(\lambda_1, \lambda_2, \ldots, \lambda_{N-1}, \lambda_N)$. We are interested to know how many eigenvalues of $A$ are needed to be able to reconstruct $A$ exactly, given the set of eigenvectors $x_1, x_2, \ldots, x_N$. In order words, we perturb the spectrum, the set of eigenvalues $\lambda_1, \lambda_2, \ldots, \lambda_{N-1}, \lambda_N$, of the adjacency matrix $A$ by omitting the $j$ smallest eigenvalues in absolute value. Since $\sum_{j=0}^{N} \lambda_j = 0$, on average, half of the eigenvalues of the adjacency matrix $A$ are negative. Therefore, we reorder the eigenvalues as $|\lambda_{(1)}| \leq |\lambda_{(2)}| \leq \cdots \leq |\lambda_{(n)}|$ such that $\lambda_{(j)}$ is the $j$th smallest (in absolute value) eigenvalue corresponding to the eigenvector $x_{(j)}$. Let us define the $N \times N$ matrices

$$\Lambda_{(j)} = \text{diag}(0, \ldots, 0, \lambda_{(j+1)}, \lambda_{(j+2)}, \ldots, \lambda_{(N)})$$

and

$$A_{(j)} = \widetilde{X} \Lambda_{(j)} \widetilde{X}^T,$$

where $\widetilde{X} = [x_{(1)} \; x_{(2)} \; \cdots \; x_{(N)}]$ is the reordered version of $X$ corresponding to the eigenvalues ranked in absolute value. Thus, $\Lambda_{(j)}$ is the diagonal matrix where the $j$ smallest (in absolute value) eigenvalues are put equal to zero or, equivalently, they are removed from the spectrum of $A$. The spectral perturbation here considered consists of consecutively removing more eigenvalues from the spectrum until we can no longer reconstruct the adjacency matrix $A$. Clearly, when $j = 0$, we have that $A_{(0)} = A$ and that, for any other $j > 0$, $A_{(j)} \neq A$. Moreover, when $j > 0$, $A_{(j)}$ is not a zero-one matrix. In Fig. 1, we show the histograms of the entries of $A_{(5)}$, $A_{(10)}$, $A_{(15)}$, and $A_{(20)}$ for an Erdős-Rényi random graph with $N = 36$ nodes and link density of $p = 0.5$. The removal of a part of the eigenvalues impacts the distribution of entries around 1 and the distribution of entries around 0 similarly, as shown in Fig. 1. This means that the deviation of entries around 1 and the deviation of entries around 0 are almost the same and that the distribution of values around 1 and 0 will reach 1/2 roughly simultaneously when the number of removed eigenvalues increases gradually. Using Heavyside's step function $h(x)$,

$$h(x) = \begin{cases} 0 & \text{if } x < 0 \\ \frac{1}{2} & \text{if } x = 0 \\ 1 & \text{if } x > 0, \end{cases}$$

we truncate the elements of $A_{(j)}$ as $h[(A_{(j)})_{ij} - \frac{1}{2}]$. If we now define the operator $\mathcal{H}$ applied to a matrix $A_{(j)}$ that replaces each element of $A_{(j)}$ by $h[(A_{(j)})_{ij} - \frac{1}{2}]$, then

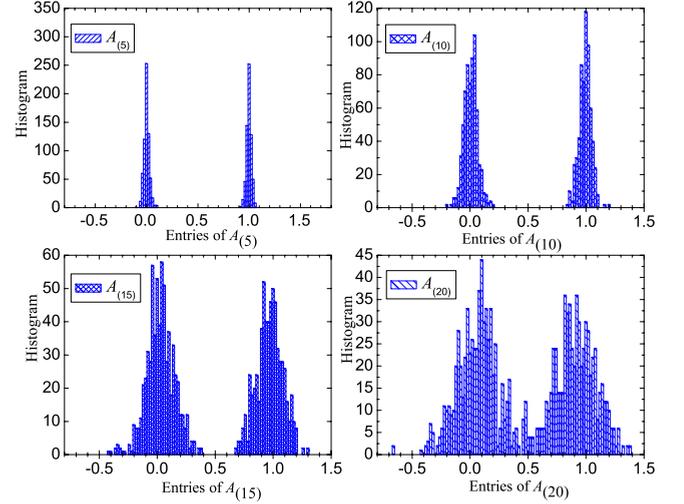

FIG. 1. (Color online) Histograms of the entries of $A_{(5)}$, $A_{(10)}$, $A_{(15)}$, and $A_{(20)}$. The matrix $A(A = A_{(0)})$ is the adjacency matrix of an Erdős-Rényi random graph with 36 nodes and link density of 0.5.

$$\widetilde{A}_j = \mathcal{H}(A_{(j)})$$

is a zero-one matrix, with the possible exception of elements $\frac{1}{2}$. The interesting observation from extensive simulation is that there seems to exist a maximal number $\theta$, such that

$$\widetilde{A}_j = A \quad \text{if } j \leq \theta,$$

$$\widetilde{A}_j \neq A \quad \text{if } j > \theta.$$

In other words, $\theta$ is the maximum number of eigenvalues that can be removed from the spectrum of the graph such that the graph can still be reconstructed precisely given the matrix $X$. We therefore call $\theta$ the *reconstructability coefficient*.

### B. Theory

The eigenvalue decomposition $A = X\Lambda X^T$ of a symmetric matrix can be rewritten in vector notation as

$$A = \sum_{k=1}^{N} \lambda_k x_k x_k^T = \sum_{k=1}^{N} \lambda_k E_k, \tag{2}$$

where the matrix $E_k = x_k x_k^T$ is the outer product of $x_k$ by itself. Any element of $A$ can be written, with the above relabeling of the eigenvectors according to a ranking in absolute values of the eigenvalues $|\lambda_{(1)}| \leq |\lambda_{(2)}| \leq \cdots \leq |\lambda_{(N)}|$, as

$$a_{ij} = \sum_{k=1}^{m} \lambda_{(k)} (E_{(k)})_{ij} + \sum_{k=m+1}^{N} \lambda_{(k)} (E_{(k)})_{ij}, \tag{3}$$

where $m \in [1, N]$ is, for the time being, an integer. As shown in Appendix A, the 2-norm of $E_k$ is not larger than 1, so that $|(E_{(k)})_{ij}| \leq 1$ for any $1 \leq k \leq N$, which implies that $-1 \leq (E_{(k)})_{ij} \leq 1$. Relation (2) also explains why an ordering in absolute value is most appropriate for our spectral perturbation: the usual ordering $\lambda_1 \geq \lambda_2 \geq \cdots \geq \lambda_{N-1} \geq \lambda_N$ in algebraic graph theory would first remove $\lambda_N < 0$, then $\lambda_{N-1}$, and so on. However, $|\lambda_N|$ can be large and its omission from the spectrum is likely to cause too big an impact.

The reconstructability of a graph as described in Sec. II A is now reformulated as follows. Since $a_{ij}$ is either 0 or 1, it follows from Eq. (3) that if

$$\left| a_{ij} - \sum_{k=m+1}^{N} \lambda_{(k)}(E_{(k)})_{ij} \right| < \frac{1}{2}, \qquad (4)$$

we can reconstruct the element $a_{ij}$ as

$$a_{ij} = \begin{cases} 1 & \text{if } \displaystyle\sum_{k=m+1}^{N} \lambda_{(k)}(E_{(k)})_{ij} > \frac{1}{2} \\ 0 & \text{if } \displaystyle\sum_{k=m+1}^{N} \lambda_{(k)}(E_{(k)})_{ij} < \frac{1}{2}. \end{cases}$$

The reconstructability requirement (4) determines the values of $m$ that satisfy the inequality. The largest value of $m$ obeying (4) is denoted by $\theta$, called the reconstructability coefficient of a graph.

Using Eq. (3), the reconstructability requirement (4) is equivalent to

$$\left| \sum_{k=1}^{\theta} \lambda_{(k)}(E_{(k)})_{ij} \right| < \frac{1}{2}.$$

A further analysis is difficult due to the appearance of the matrix elements $(E_{(k)})_{ij}$ of which, in general, is not much known. Since $|(E_{(k)})_{ij}| \le 1$, we can bound the sum as

$$\left| \sum_{k=1}^{\theta} \lambda_{(k)}(E_{(k)})_{ij} \right| \le \sum_{k=1}^{\theta} |\lambda_{(k)}||(E_{(k)})_{ij}| \le \sum_{k=1}^{\theta} |\lambda_{(k)}|. \qquad (5)$$

In many cases, this bound is conservative because, on average, half of the eigenvalues of the adjacency matrix $A$ are negative. Moreover, the matrix element $(E_{(k)})_{ij}$ can also be negative. We show in Appendix B for the class of Erdős-Rényi random graph $G_p(N)$ that the bound (5) is, indeed, too conservative and that only extensive simulations seem appropriate to determine the reconstructability coefficient $\theta$.

## III. RECONSTRUCTABILITY COEFFICIENT OF ERDŐS-RÉNYI RANDOM GRAPHS $G_p(N)$

Traditionally, complex networks have been modeled as Erdős-Rényi (ER) random graphs $G_p(N)$, which can be generated from a set of $N$ nodes by randomly assigning a link with probability $p$ to each pair of nodes. Besides their analytic tractability [9], the ER random graphs have also served as idealized structures for peer-to-peer networks, *ad hoc* networks, gene networks, ecosystems, and the spread of diseases or computer viruses. If a graph problem cannot be solved analytically for $G_p(N)$, experience teaches that chances are high that the problem is analytically intractable for all graphs with at least one parameter that can be changed (such as $N$). For this reason, ample attention is devoted to analyze the behavior of the reconstructability coefficient $\theta_p(N)$ for the random graph $G_p(N)$.

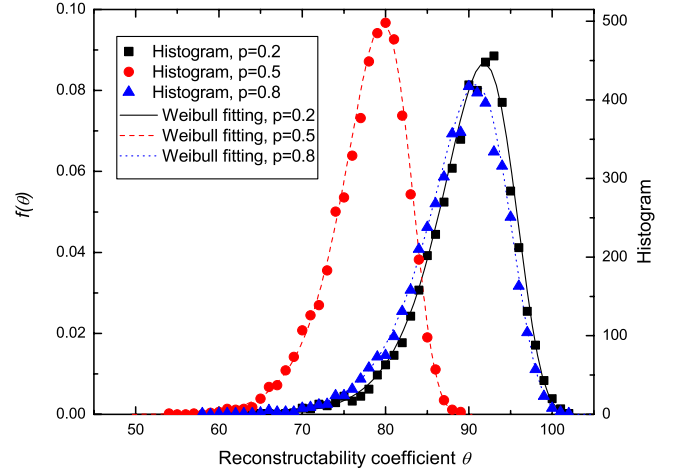

FIG. 2. (Color online) Probability density function $f(\theta)$ for $N = 200$ and $p = 0.2, 0.5, 0.8$. For each $p$, 5000 samples of $\theta$ are computed and the histograms are fitted by Weibull distribution.

According to the Wigner's semicircle law [10] and the fact that the complement of $[G_p(N)]^c = G_{1-p}(N)$, for sufficiently large $N$, the spectrum of $G_p(N)$ is symmetric around $p = \frac{1}{2}$. Therefore, we expect that $\theta_p(N) = \theta_{1-p}(N)$ for sufficiently large $N$, which is confirmed by simulations below.

### A. Weibullian probability distribution of $\theta_p(N)$

We explore three classes of ER random graphs: $G_{0.2}(200)$, $G_{0.5}(200)$, and $G_{0.8}(200)$. For each class, 5000 random graphs are generated and we compute the reconstructability coefficient $\theta_p(N)$ of each graph. Figure 2 illustrates the probability distribution of the reconstructability coefficient $\theta_p(N)$ of ER random graphs, fitted by a Weibull distribution. The probability density function of Weibull distribution is

$$f_{\text{Weibull}}(x; \phi, \varphi) = \begin{cases} \dfrac{\varphi}{\phi}\left(\dfrac{x}{\phi}\right)^{\varphi-1} \exp\left[ -\left(\dfrac{x}{\phi}\right)^{\varphi} \right] & \text{if } x \ge 0 \\ 0 & \text{if } x < 0, \end{cases}$$

where $\phi$ is the scale parameter and $\varphi$ is the shape parameter [11].

### B. $E[\theta_p(N)]$ as a function of $N$ and $p$

Here, we investigate the average reconstructability coefficient $E[\theta_p(N)]$ as a function of $N$ and $p$ for ER random graphs. First, we consider the ER random graphs $G_{0.5}(N)$, where the graph size $N$ is increased from $N = 50$ to $N = 2500$ with a step of 50. Within each class $G_{0.5}(N)$, the realizations of $\theta$ for 10 000 graphs are computed. The mean $E[\theta]$ as well as the standard deviation of $\theta$ are depicted in Fig. 3. $E[\theta]$ appears to be linear with $N$ as fitted in Fig. 3. The linear fitting function of $E[\theta]$ as a function of $N$, for $G_{0.5}(N)$, is

$$E[\theta_{0.5}(N)] = 0.36N.$$

Furthermore, Fig. 4 shows the scaling law of $E[\theta_p(N)]$ as a function of $N$ for random graphs with link density $p$ from 0.1 to 0.9. These extensive simulations suggest the linear scaling law (1). The magnitude of $a$ for the ER random graphs is surprisingly large. In $G_{0.5}(N)$, about 40% of the spectrum is redundant from the point of view of reconstructability defined above. For $G_p(N)$ with $p=0.1, 0.2, \ldots, 0.9$, the parameters $a$ of the general linear scaling law (1) are shown in Table I. Thus, we further examine the slope $a(p)$ as a function of the link density $p$ in Fig. 5. From Table I, we observe the approximate symmetry of the slope $a(p)$. The slope $a(p)$ as a function of $p$ is fitted with a parabola

$$a(p) = 0.39(p - 0.5)^2, \qquad (6)$$

as shown in Fig. 5.

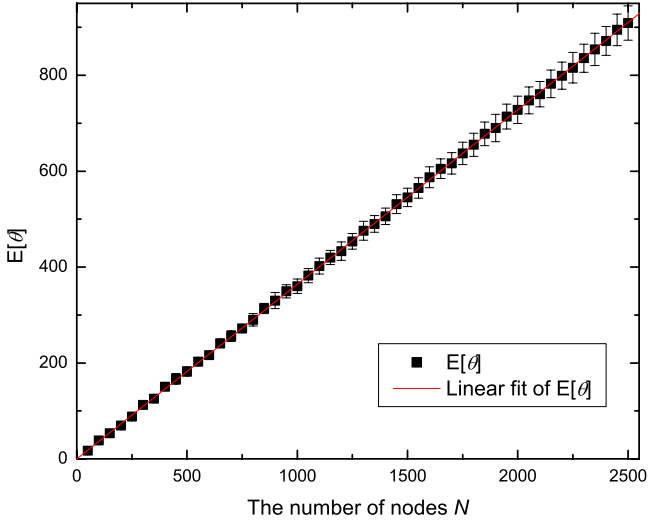

FIG. 3. (Color online) Mean of reconstructability coefficient $E[\theta]$ of ER random graphs $G_{0.5}(N)$ with standard deviation $\sigma_\theta$ shown as Y-error bar. $E[\theta]$, which is fitted linearly in the figure, is computed from 10 000 samples for each $N$.

Let $E[\theta(p,N)]$ be the function $E[\theta]$ as a function of $p$ when there are $N$ number of nodes. We consider the ER

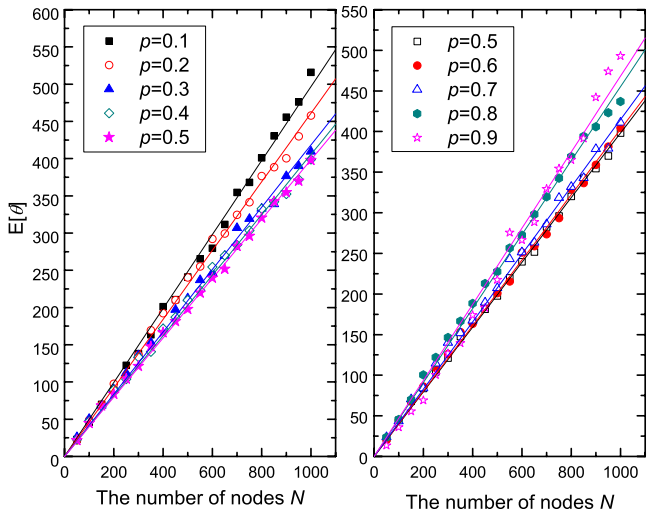

FIG. 4. (Color online) Scaling law for random graphs with $p = 0.1, 0.2, \ldots, 0.9$. We compute 10 000 samples of $\theta$ to get the mean $E[\theta]$.

TABLE I. The parameter $a$ of the general scaling law for ER random graphs.

| $p$ | 0.1 | 0.2 | 0.3 | 0.4 | 0.5 | 0.6 | 0.7 | 0.8 | 0.9 |
|---|---|---|---|---|---|---|---|---|---|
| $a$ | 0.50 | 0.46 | 0.42 | 0.40 | 0.39 | 0.40 | 0.41 | 0.45 | 0.47 |

random graphs $G_p(200)$ with $N=200$ nodes and $p$ ranging from $p=0.01$ to 1 with a step of 0.01. Similarly, 10 000 graphs are generated for each class $G_p(200)$. As depicted in Fig. 6, the $E[\theta]$ as a function of the link density $p$ follows a similar parabola

$$E[\theta(p,200)] = 77(p - 0.5)^2, \qquad (7)$$

where $E[\theta]$ appears to be symmetric around $p=0.5$ within the range $p \in [0.09, 0.91]$. It seems that, when $N \to +\infty$, $E[\theta(p,200)] = 77(p - 0.5)^2$ holds generally for the ER random graphs. The parabola fitting only works well in the region $p \in [0.09, 0.91]$, where the graph is well connected, and so is its complementary graph, which means that the graph is not too sparse and not too dense. For very sparse or dense graphs, the analysis of the reconstructability coefficient is complicated because many degenerate eigenvalues appear.

## IV. RECONSTRUCTABILITY COEFFICIENT OF OTHER TYPES OF NETWORKS

In this section, we will examine whether the linear scaling law (1) between the average reconstructability coefficient $E[\theta]$ and the size $N$ of a network is generally true for other types of networks.

### A. Scale-free networks

Power law graphs are random graphs specified by a power law degree distribution $\Pr[D=k] = L(k)k^{-\tau}$, where $L(k)$ is a slowly varying function of $k$ [12]. The power law degree distribution is followed by many natural and artificial networks such as the scientific collaborations, the world-wide web, and the internet. Specifically, we investigate the Barabási-Albert power law graph [13–15], which starts with $m_0$ nodes. At every time step, we add a new node with $m$ links that connect the new node to $m$ different nodes already present in the graph. The probability that a new node will be connected to node $i$ in step $t$ is proportional to the degree $d_i(t)$ of that node.

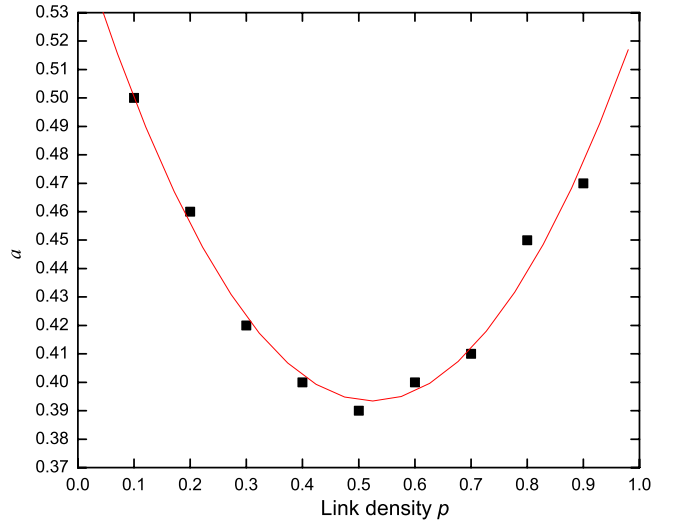

FIG. 5. (Color online) Curve fitting of the slope $a(p)$.

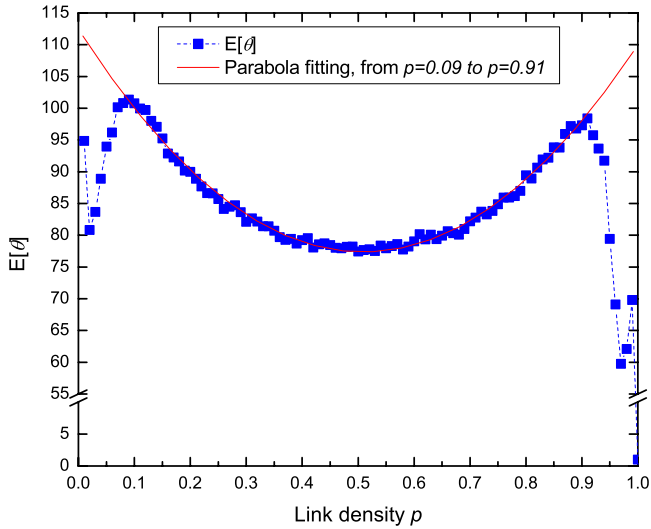

FIG. 6. (Color online) $E[\theta]$ of Erdős-Rényi random graphs $G_p(200)$ for $p$. $E[\theta]$ is computed by 10 000 samples of $\theta$. The function $E[\theta]$ of $p$, where $p \in [0.09, 0.91]$, is fitted by a parabola.

We consider the Barabási-Albert graphs with $N$ ranging from 500 to 2500 with a step of 100 and $m = 3, 5, 7$. Large network sizes $N$ are selected because a power law degree distribution can be observed only when the network size is large. Within each class of the BA graphs with a specific $N$ and $m$, 10 000 graphs are generated. In Fig. 7, we observe that the linear scaling law (1) seems to hold for Barabási-Albert networks as well. The slopes $a$ of the corresponding linear fitting are shown in Table II.

TABLE II. The parameter $a$ of the general scaling law for scale-free networks.

| $m$ | 3 | 5 | 7 |
|---|---|---|---|
| $a$ | 0.39 | 0.42 | 0.46 |

### B. Small-world networks

The small-world model proposed by Watts and Strogatz [16] encompasses the following two structural features as observed in real-world networks. Any two nodes can be reached within a small number of links despite the large size of networks. Nodes are well clustered in the sense that two direct neighbors of a node are more likely to be connected compared to those in random graphs. The small-world model starts by building a ring with $N$ nodes and by joining each node with $k$ nearest neighbors ($k/2$ on either side of the ring). Upon the resulted ring lattice, each link connected to a clockwise neighbor is rewired to a randomly chosen node with a probability $p_r$ and is preserved with a probability $1 - p_r$. The small-world graph interpolates between a ring or lattice ($p_r = 0$) and a random graph with the constraint that each node has the minimal degree $k/2$ ($p_r = 1$).

The linear scaling law (1) of $E[\theta]$ as a function of $N$ is also observed in Watts-Strogatz networks with different $k$ and $p_r$, as shown in Fig. 8. Table III shows the parameters of the linear scaling law obtained from the curve fitting in Fig. 8. The slope $a$, or the proportion of eigenvalues that can be removed while the adjacency matrix is still reconstructable, depends on the average degree $k$ and the rewiring probability $p_r$.

### C. Deterministic graphs

Finally, we explore the average reconstructability coefficient $E[\theta]$ as a function of $N$ in a set of deterministic graphs:

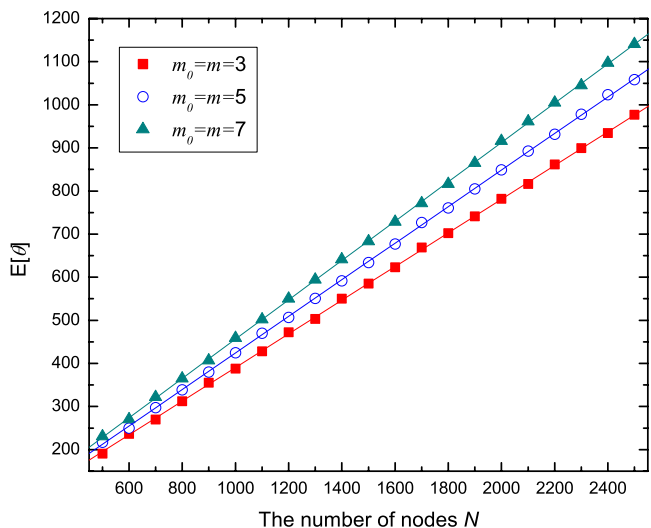

FIG. 7. (Color online) Linear scaling law for $E[\theta]$ of Barabási-Albert networks. $E[\theta]$ is obtained by 10 000 samples. $m_0$ is the number of nodes of the initial graph and for each time step, the coming node can add $m$ links to the current graph.

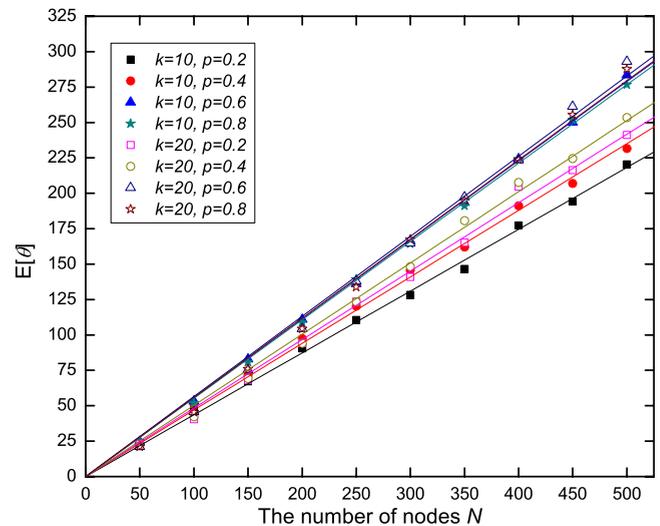

FIG. 8. (Color online) Linear scaling law for $E[\theta]$ of Watts-Strogatz networks. $E[\theta]$ is obtained by 10 000 samples. For the initial ring lattice graph, every node has links with its $k$ nearest nodes and $p_r$, defined as the randomness, is the probability with which each link is rewired.

TABLE III. The parameter $a$ of the general scaling law for small-world networks.

| $k$ | 10 | 10 | 10 | 10 | 20 | 20 | 20 | 20 |
|---|---|---|---|---|---|---|---|---|
| $p$ | 0.2 | 0.4 | 0.6 | 0.8 | 0.2 | 0.4 | 0.6 | 0.8 |
| $a$ | 0.44 | 0.47 | 0.56 | 0.55 | 0.48 | 0.50 | 0.57 | 0.56 |

TABLE IV. The parameter $a$ of the general scaling law for deterministic graphs.

| Graph type | Path | Cycle | Wheel | Grid | Cube | Four-dimensional lattice |
|---|---|---|---|---|---|---|
| $a$ | 0.73 | 0.73 | 0.73 | 0.67 | 0.73 | 0.76 |

(a) a path; (b) a ring where each node on a circle is connected to its previous and subsequent neighbor on the ring; (c) a wheel where a node locates in the wheel center while the other nodes are on a circle around the wheel center and the wheel center is connected to every node on the outer circle while each node on the outer circle connects to its previous and subsequent neighbor. (d) $D$-lattice or $D$-dimensional lattices where all interior nodes have the same degree and $D$ is the dimension. For the first three types of graphs, we increase the number of nodes $N$ from 100 to 800 with a step of 100. Here, we confine ourselves to the hypercube $D$ lattices in which each edge is of equal size. In this case, a 2D lattice becomes a grid and a three-dimensional (3D) lattice equals a cubic lattice. For grid graphs, we take $N=(5k)^2$, $k=1,2,\ldots,6$, for cubic graphs, $N=k^3$, $k=2,3,\ldots,9$. The 4D lattices with $N=k^4$, where $k=2,3,4,5$ are considered. Figure 9 shows the reconstructability coefficient $\theta$ of each graph, which seems always a linear function of the network size $N$. The corresponding linear curve fittings are described in Table IV. The paths, ring graphs, and wheel graphs follows a similar linear relation between $\theta$ and $N$ because of similarity among their topologies. As the dimension $D$ of a lattice increases, the slope $a$ is larger. In other words, more eigenvalues can be removed without influencing the reconstruction of the adjacency matrix.

The complete bipartite graph $K_{m,n}$ consists of two sets $\mathcal{M}$ and $\mathcal{N}$ with $m=|\mathcal{M}|$ and $n=|\mathcal{N}|$ nodes, respectively, where each node of one set is connected to all other nodes of the other set. There are no connections between nodes of a same set. The eigenvalues of the adjacency matrix $A_{K_{nm}}$ are $\lambda_1=-\lambda_N=\sqrt{mn}=\sqrt{m(N-m)}$ and all others are zero. Hence, only two eigenvalues contribute in Eq. (2) such that $\theta \gtrsim N-2$. For large $N$, the general observed law (1) then shows that $a_{K_{nm}}=\frac{E[\theta_{K_{nm}}]}{N}\to 1$. In other words, the maximum possible limit value of $a$ can be attained in very large complete bipartite graphs. However, this example shows that some graphs may have intrinsic zero eigenvalues, which should be distinguished from zero eigenvalues introduced by perturbations. Fortunately, for large complex networks, the probability that intrinsic zero eigenvalues occur tends to zero, such that the reconstructability coefficient $\theta$ mostly measures a resilient property of a large complex networks against a spectral perturbation.

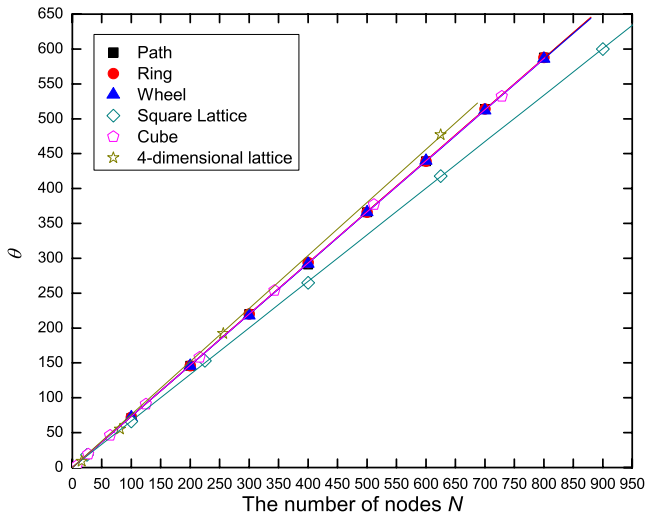

FIG. 9. (Color online) Linear scaling law for $E[\theta]$ of six special deterministic types of graphs. For path graph, ring graph, and wheel graphs, $N=100,200,300,\ldots,800$. We take $N=25,100,225,400,625,900$, $N=8,27,64,125,216,343,512,729$, and $N=16,81,256,625$ for grid graphs, cubic graphs, and four-dimensional lattice graphs, respectively.

### D. Summary

Surprisingly, a linear scaling law between the average reconstructability $E[\theta]$ and the network size $N$ has been observed in ER random graphs, power law graphs, small-world graphs, and various deterministic graphs. This suggests that the linear relation $E[\theta]=aN$ may be a generic feature possessed by various complex networks.

## V. LINEAR SCALING LAW WITH EIGENVECTOR PERTURBATION FOR ERDŐS-RÉNYI RANDOM GRAPH

Besides the perturbation on eigenvalues, the eigenvector matrix can be perturbed. Recall that $X=[x_1 \ x_2 \ \cdots \ x_N]$ denotes the eigenvector matrix of the graph with $N$ nodes. We generate a perturbation matrix $R$, which is an $N$ by $N$ matrix with random entries, chosen independently from a normal distribution with mean of 0 and variance of 1.

The Euclidean norm of a square matrix is identical to its largest singular value. Since $\|X\|_2=1$ for all the graph sizes $N$, we normalize the perturbation matrix as $\bar{R}=\frac{R}{2\cdot\sqrt{N}}$ such that the norm of the new perturbation matrix $\|\bar{R}\|_2\approx 1$ for all the graph sizes $N$. The perturbed eigenvector matrix is defined as

$$X'=X+\varepsilon\bar{R}, \qquad (8)$$

where the constant $\varepsilon$ is called the perturbation factor. Under such perturbation on the eigenvector matrix, we re-examine the reconstructability coefficient.

In this section, we investigate the linear scaling law with normalized eigenvector perturbation defined by Eq. (8), confined to the ER random graphs. Intuitively, with strong perturbation in the eigenvector matrix (large $\varepsilon$), the reconstruc-tability coefficient is expected to be small. Figure 10 shows the linear scaling law for Erdős-Rényi random graphs. With normalized eigenvector perturbation, the linear scaling law (1) still holds for Erdős-Rényi random graphs when $\varepsilon \leq 0.2$ (Table V).

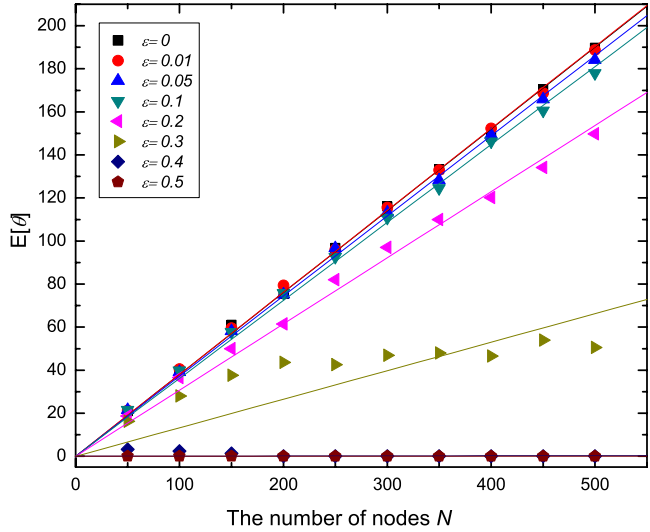

FIG. 10. (Color online) Linear scaling law with normalized eigenvector perturbation for Erdős-Rényi random graphs. $E[\theta]$ is the mean of 10 000 samples of $\theta$. The scaling factors $\varepsilon$ are shown in the legend. The link density $p=0.5$.

## VI. CONCLUSION

In this paper, we have studied the spectral reconstructability of complex networks and defined a metric of networks: the reconstructability coefficient $\theta$. Through extensive simulations, we investigated the properties of the reconstructability coefficient $\theta$ for several important types of complex networks, such as ER random graphs, scale-free networks, and small-world networks, and also other special deterministic types of graphs. A general linear scaling law (1), $E[\theta]=aN$, was found. For sufficiently large $N$, a portion $a$ of the smallest eigenvalues (in absolute value) can be removed from the spectrum and the adjacency matrix is still reconstructable with its original eigenvectors. The magnitude of $a$ for different types of complex networks with different parameters varies from 39% to 76%, as shown in Table VI.

The properties of the mean of the reconstructability coefficient $E[\theta]$ were also studied under eigenvector perturbation for ER random graphs. The normalized Gaussian distributed noise matrix, scaled by the perturbation factor $\varepsilon$, was added to the eigenvector matrix $X$. Simulations show that the linear scaling law $E[\theta]=aN$ still holds for ER random graphs until the perturbation factor $\varepsilon$ exceeds 0.2.

TABLE V. The parameter $a$ of the general scaling law for ER random graphs under eigenvector perturbation.

| $\varepsilon$ | 0 | 0.01 | 0.05 | 0.1 | 0.2 | 0.3 | 0.4 | 0.5 |
|---|---|---|---|---|---|---|---|---|
| $a$ | 0.38 | 0.38 | 0.37 | 0.36 | 0.31 | 0.13 | 0 | 0 |

TABLE VI. The summary of the parameter $a$ for different types of graphs or networks.

| | $a$ | |
|---|---|---|
| | min | max |
| ER random graphs | 0.39 | 0.5 |
| Barabási-Albert power law graphs | 0.39 | 0.46 |
| Watts-Strogatz small-world graphs | 0.44 | 0.57 |
| Deterministic graphs | 0.67 | 0.76 |

The basic eigenvalue relation (2) shows that the sets of orthogonal eigenvectors are weighted by their corresponding eigenvalues. Any eigenvector specifies an orthogonal direction in the $N$-dimensional space. The eigenvector with an eigenvalue in absolute value close to zero contains redundant information about the topology of the graph, in the sense that after the removal of this eigenvalue, the network can still be reconstructed from the remaining spectrum. We observe that when the graphs have more constraints to be generated, the parameter $a$ is higher. Those deterministic graphs, such as path, ring, and grid graphs, have more constraints to be generated than ER random graphs, power law graphs, and small-world graphs. In the spectral domain, the more generating constraints the graphs have, the less that $N$-dim space is "sampled," or, in other words, the less spectral bases (eigenvectors) we need to reconstruct the graphs. One may also say that the embedding of the graph structure in the $N$-dim space does not need those orthogonal dimensions (that act similarly as a kernel of a linear transformation).

## APPENDIX A: PROPERTIES OF THE MATRIX $E_k$.

From the definition $E_k = x_k x_k^T$, we deduce that $E_k = E_k^T$, thus, symmetric. The explicit form of the matrix $E_k$ is

$$E_k = x_k x_k^T = \begin{bmatrix} (x_{k1})^2 & x_{k1}x_{k2} & x_{k1}x_{k3} & \cdots & x_{k1}x_{kn} \\ x_{k2}x_{k1} & (x_{k2})^2 & x_{k2}x_{k3} & \cdots & x_{k2}x_{kn} \\ x_{k3}x_{k1} & x_{k3}x_{k2} & (x_{k3})^2 & \cdots & x_{k3}x_{kn} \\ \vdots & \vdots & \vdots & \vdots & \vdots \\ x_{kn}x_{k1} & x_{kn}x_{k2} & x_{kn}x_{k3} & \cdots & (x_{kn})^2 \end{bmatrix},$$

which shows that the diagonal element $(E_k)_{ii}=(x_{ki})^2$ equals the square of the $i$th vector component of the eigenvector $x_k$. Hence,

$$\mathrm{tr}(E_k) = \sum_{i=1}^{n} (x_{ki})^2 = x_k^T x_k = 1. \tag{A1}$$

It follows from the orthogonality property (A21) in [3] of eigenvectors $x_k$ of a symmetric matrix that $E_k^2=E_k$ and $E_k E_m=0$ for $k \neq m$. Let us denote the eigenvalue equation $E_k y_j = \xi_j y_j$ of the symmetric matrix $E_k$. After left multiplication by $E_k$, we obtain $E_k^2 y_j = \xi_j^2 y_j$ and, since $E_k^2=E_k$, we arrive at $E_k y_j = \xi_j^2 y_j$. Hence, for any eigenvalue $\xi_j$ and corresponding eigenvector $y_j$, we have that $\xi_j y_j = \xi_j^2 y_j$, which implies that $\xi_j$ is either zero or 1. The trace relation (A7) in [3] and Eq. (A1) indicates that $\sum_{j=1}^{n} \xi_j = 1$. The eigenvalues of $E_k$ directly follow from the rank-one update formula because

$$\det(x_k x_x^T - \lambda I) = (-\lambda)^n \det\left(I - \frac{1}{\lambda} x_k x_x^T\right) = (-\lambda)^{n-1}(\lambda - 1) \tag{A2}$$

and are precisely the same as those of the adjacency matrix of the complete graph $K_n$. Consequently, we conclude that $n-1$ eigenvalues are zero and one eigenvalue equals 1, such that $\|E_k\|_2 = 1$ that follows from (A33) in [3]. The zero eigenvalues imply that $\det(E_k) = 0$ and that the inverse of $E_k$ does not exist. Geometrically, this is understood because, by projecting, information is lost and the inverse cannot create information.

## APPENDIX B: QUALITY OF THE BOUND (5)

When the graph belongs to the class of Erdős-Rényi random graph $G_p(N)$, the spectrum rapidly tends to Wigner's semicircle law [3]. We confine ourselves to the class of Erdős-Rényi random graph $G_p(N)$ to estimate the quality of the bound (5). In particular, we compute the bound (5) of $\theta$ probabilistically as

$$\Pr\left[\sum_{k=1}^{\theta} |\lambda_{(k)}| < \frac{1}{2}\right] = 1 - \epsilon,$$

meaning that the probability that $\sum_{k=1}^{\theta} |\lambda_{(k)}| < \frac{1}{2}$ is almost sure, when $\epsilon > 0$ is chosen arbitrarily small. However, the distribution of the $\theta$ smallest order statistics is difficult and we content ourselves to compute the average of the sum of order statistics

$$r = \sum_{k=1}^{\theta} E[|\lambda_{(k)}|].$$

First, we compute the absolute value $Y$ of a random variable $X$. The event that $\{Y \le y\}$ is equivalent to $\{|X| \le y\} = \{-y \le X \le y\}$ and nonexistent for $y < 0$. Hence,

$$\Pr[Y \le y] = \Pr[-y \le X \le y] = F_X(y) - F_X(-y)$$

and, after differentiation with respect to $y$, we find the relation for the probability density function as

$$f_{|X|}(y) = f_X(y) + f_X(-y).$$

Applied to Wigner's semicircle law,

$$f_{|\lambda(A/\sqrt{N})|}(x) = \frac{1}{\pi \sigma^2} \sqrt{4\sigma^2 - x^2} \, 1_{|x| \le 2\sigma}$$

and we find the distribution $F_{|\lambda(A/\sqrt{N})|}(t) = \frac{1}{\pi \sigma^2} \int_0^t \sqrt{4\sigma^2 - x^2} \, 1_{|x| \le 2\sigma} dx$ for $t \le 2\sigma$ as

$$F_{|\lambda(A/\sqrt{N})|}(t) = \Pr\left[\left|\lambda\left(\frac{A}{\sqrt{N}}\right)\right| \le t\right]$$

$$= \frac{2}{\pi} \arcsin \frac{t}{2\sigma} + \frac{2}{\pi} \frac{t}{2\sigma} \sqrt{1 - \left(\frac{t}{2\sigma}\right)^2},$$

while $F_{|\lambda(A/\sqrt{N})|}(t) = 1$ for $t > 2\sigma$. The Taylor expansion is

$$F_{|\lambda(A/\sqrt{N})|}(2\sigma x) = \frac{4}{\pi} x \left(1 - \frac{1}{6} x^2 + O(x^4)\right),$$

which shows that, for small $x$,

$$F_{|\lambda(A/\sqrt{N})|}(2\sigma x) \simeq \frac{4}{\pi} x. \tag{B1}$$

Second, using a similar argument as in [3] that approximates the exact distribution of the $k$th order statistics as a Gaussian with mean

$$\mu = N F_{\lambda(A/\sqrt{N})}(x)$$

and variance

$$\sigma^2 = N F_{\lambda(A/\sqrt{N})}(x)(1 - F_{\lambda(A/\sqrt{N})}(x))$$

that tends to a delta function for large $N$, we can approximate

$$E[\lambda_{(k)}] \simeq 2\sigma \sqrt{N} F_{\lambda(A/\sqrt{N})}^{-1}\left(\frac{k}{N}\right).$$

Using Eq. (B1) yields

$$E[\lambda_{(k)}] \simeq \sqrt{N} \pi \sigma^2 \left(\frac{k}{N}\right) = \frac{\pi \sigma^2}{\sqrt{N}} k,$$

such that

$$r = \sum_{k=1}^{\theta} E[|\lambda_{(k)}|] \simeq \frac{\pi \sigma^2}{\sqrt{N}} \sum_{k=1}^{\theta} k = \frac{\pi \sigma^2}{2\sqrt{N}} \theta(\theta + 1) \simeq \frac{\pi \sigma^2}{2\sqrt{N}} \theta^2.$$

The requirement that $r < \frac{1}{2}$ then implies approximately that

$$\theta < \frac{N^{1/4}}{\sigma \sqrt{\pi}}.$$

The derivation shows that the conservative bound is inappropriate because simulations show that $\theta = O(N)$, while the conservative bound points to $\theta = O(N^{1/4})$.